\documentstyle[prc,multicol,aps]{revtex}
\author{J.~H.~Morrison,\footnote{Present Address:  University
	of Houston Law Center, Houston, TX 77204}
H.~Baghaei,\footnote{Present address:
	Dept of Nuclear Medicine, UT MD Anderson Cancer Center,
	1100 Holcombe Blvd, Houston, TX 77030}
W.~Bertozzi, S. Gilad, J.~Glickman,\footnote{Present Address: 9308 Cheney
	Hill Rd., College Park, MD 20740}
C.~E.~Hyde-Wright,\footnote{Present Address:  Department of Physics,
	Old Dominion University, Norfolk, VA 23529}
\edef\ODUfootnote{\arabic{footnote}}
N.~Kalantar-Nayestanaki,\footnote{Present address: KVI, 9747 AA
	Groningen, The Netherlands}
R.~W.~Lourie,\footnote{Present address: SUNY at Stony Brook, Stony Brook,
	NY 11794-3800}
S.~Penn,\footnote{Present address: Department of Physics, Syracuse University,
	Syracuse, NY, 13244}
P.~E.~Ulmer,\footnotemark[\ODUfootnote]
L.~B.~Weinstein\footnotemark[\ODUfootnote]}
\address{Department of Physics, Massachusetts Institute of Technology,
	Cambridge, MA 02139}
\author{B.~H.~Cottman,\footnote{Present address: I-Kinetics, Inc., 19
	Bishop Allen Drive, Cambridge, MA 02139}
L.~Ghedira, E.~J.~Winhold}
\address{Department of Physics, Rensselaer Polytechnic Institute,
	Troy, New York 12181}
\author{J.~R.~Calarco, J.~Wise\footnote{Present address: 74 Boston Rd.
	Apt. B-106, Chelmsford, MA 01824}}
\address{Department of Physics, University of New Hampshire,
	Durham, NH 03824}
\author{P.~Boberg,\footnote{Present Address:  USRA, Code 7654, Naval
	Research Laboratory, Washington, DC 20375-5352}
C.~C.~Chang, D.~Zhang}
\address{Department of Physics and Astronomy, University of Maryland,
	College Park, Maryland 20742}
\author{K.~Aniol, M.~B.~Epstein, D.~J.~Margaziotis}
\address{Department of Physics and Astronomy, California State
	University, Los Angeles, California, 90032}
\author{J.~M.~Finn, C.~Perdrisat, V.~Punjabi\footnote{Present Address:
	Department of Physics, Norfolk State University, Norfolk, VA}}
\address{Department of Physics, College of William and Mary,
	Williamsburg, Virginia 23185}
\title{Quasielastic $^{12}$C(e,e$'$p) Reaction at High Momentum
Transfer}
\begin{document}
\draft
%
%
%
%\twocolumn[\hsize\textwidth\columnwidth\hsize\csname@twocolumnfalse%
%\endcsname
\maketitle
\begin{abstract}
We measured the $^{12}$C(e,e$'$p) cross section as a function of
missing energy in parallel kinematics for $(q,\omega)$ = (970 MeV/c,
330 MeV) and (990 MeV/c, 475 MeV).  At $\omega=475$ MeV, at the
maximum of the quasielastic peak, there is a large continuum ($E_m > 50$ MeV) cross
section extending out to the deepest missing energy
measured, amounting to almost 50\% of the measured cross section.  The
ratio of data to DWIA calculation is 0.4 for both the p- and s-shells.
At $\omega=330$ MeV, well below the maximum of the quasielastic peak,
the continuum cross section is much smaller and the ratio of data to
DWIA calculation is 0.85 for the p-shell and 1.0 for the s-shell.  We
infer that one or more mechanisms that increase with $\omega$
transform some of the single-nucleon-knockout into multinucleon
knockout, decreasing the valence knockout cross section and increasing
the continuum cross section.
\end{abstract}
\twocolumn
\narrowtext
\section{Introduction}
This paper reports a measurement of the quasielastic $^{12}$C(e,e$'$p)
reaction at momentum transfer $q \approx 1000$ MeV/c and two energy
transfers, $\omega = 330$ MeV and $\omega = 475$ MeV.  After an
introductory discussion, we describe the experiment and its analysis.
We present a representation of the differential cross section's
$\omega$ dependence around each of the two central values, using
Legendre polynomials.  Finally, we discuss the results of the
experiment in terms of single-nucleon knockout, multinucleon knockout,
and other processes.  The PhD thesis of Morrison\cite{Morrison}
presents the experiment in more detail.

We define several quantities here: $\omega$ is the energy transferred
from the electron to the nuclear system.  The 3-momentum transfer is
$\bf q$, with magnitude $q$.  The momentum transfer four-vector is $Q
\equiv (\omega, {\bf q})$, and $Q^2 = q^2 - \omega^2$.  The missing
energy of the coincidence reaction is $E_m \equiv \omega - T_p$, where
$T_p$ is the outgoing proton's kinetic energy.  $M$ is the mass of the
nucleon.  The missing momentum is ${\bf p}_m \equiv {\bf q} - {\bf
p}_p$ where ${\bf p}_p$ is the outgoing proton's momentum.

At quasielastic kinematics, $\omega \approx Q^2/2M$, interactions with
independent nucleons are expected to dominate the nuclear
electromagnetic response.  However, despite the apparent agreement of
non-relativistic Fermi gas calculations\cite{Moniz} with quasielastic
(e,e$'$) measurements for a large range of nuclei\cite{Whitney},
measurements of the separated longitudinal and transverse (e,e$'$)
cross sections have shown that other processes contribute
significantly to the reaction.  The longitudinal and transverse
reduced response functions, $f_L$ and $f_T$, for $^3$He at $q \approx
500$ MeV/c\cite{Marchand} are equal, in accordance with the
predictions of independent particle models.  However, $f_L$ is
$\approx$40\% smaller than $f_T$ for heavier nuclei including
$^4$He, $^{12}$C, $^{40}$Ca, $^{56}$Fe, and
$^{238}$U\cite{Reden,Altemus,Finn,Meziani2,Meziani3,Barreau,Blatchley}
at $q \approx 500$ MeV/c.  This indicates the presence of a
non-quasifree process that may depend on the density or number of
available nucleons.

Yates {\em et al.\/}\cite{Yates} measured a different result on
$^{40}$Ca: $f_L$ is less than 20\% smaller than $f_T$.  At
a larger momentum transfer, $q=1050$ MeV/c, $f_L$ and $f_T$ were
comparable for both $^3$He and $^4$He on the low $\omega$ side of the
quasielastic peak\cite{Meziani}, but $f_T$ was still significantly
larger than $f_L$ at $q = 1050$ MeV/c for $^{56}$Fe\cite{Chen}. Thus
there is some experimental ambiguity in the magnitude and
momentum-transfer dependence of the transverse-longitudinal
ratio. 

Many different models of inclusive quasielastic electron scattering
attempt to treat aspects of the reaction correctly, but no model can
explain all of the data.  Such older models include $\sigma$-$\omega$
calculations\cite{Serot}, meson exchange currents\cite{vanOrden},
two-particle-two-hole models\cite{Alberico}, modification of the mass
and/or the size of the nucleon\cite{Noble,Celenza}, and quark
effects\cite{Mulders}.

Recent Green's Function monte-carlo (GFMC) calculations by Carlson and
Schiavilla\cite{Carlson}, which include pion degrees of freedom, final
state interactions, and two-body currents, can reproduce the $^3$He
and $^4$He longitudinal and transverse response functions.  They
interpret the PWIA response quenching as due to the charge-exchange
component of the nuclear interaction, which shifts the strength to
higher excitation energy.  The quenching of the transverse response is
more than offset by the contribution of two-body currents associated
with pion-exchange.  This work indicates the necessity of including
correlated initial state wave functions, two-body reaction mechanisms,
and final state interactions.  We expect that more reaction
mechanisms, including real pions, deltas, and three-nucleon currents,
need to be included for heavier nuclei and higher excitation energies.
Unfortunately, no GFMC calculations are possible yet for heavier
nuclei. 

Coincidence (e,e$'$p) electron scattering, in which a knocked out
proton is detected in coincidence with the scattered electron, can
distinguish among some of the various reaction processes proposed,
because different reactions occur at different missing energies.

The C(e,e$'$p) cross section was first measured at
Saclay\cite{Mougey} out to $E_m \approx 60$ MeV, and more recently by
van der Steenhoven\cite{Steenhoven}.  The spectrum exhibits a large
narrow peak at $E_m \approx 16$ MeV, several small, narrow peaks at
larger missing energies, and a broad structure from 25 MeV to 60 MeV.
The momentum distributions indicated that the narrow peaks correspond
to the knockout of a proton in a p-shell state, while the broad
structure results from s-shell proton knockout.  The spectroscopic
factors were reported as 2.5 for the p-shell peaks, and 1.0 for the
s-shell peak.\cite{Mougey}   The s-shell peak is broad because the residual nucleus
is in an excited state, and decays rapidly.  Two-nucleon knockout may
also contribute to the strength in the s-shell region as the threshold
for this process is at $E_m \approx 27$ MeV.

Lapik\'as\cite{Lapikas} has found the strength for valence shell
knockout in (e,e$'$p) to be reduced by 20\% for elements throughout
the periodic table.

Several experiments at Bates have measured the C(e,e$'$p) cross
section as a function of missing energy for the following kinematical
conditions: The maximum of the quasielastic peak at $q = 400$~MeV/c
(an L/T-separation), 585, 775, and 827~MeV/c\cite{Ulmer,Weinstein};
the dip region at $q = 400$~MeV/c\cite{Lourie}, and the delta peak at
$q = 400$ and 475~MeV/c\cite{Baghaei2}.  These measurements had four
major results:
\begin{enumerate}
\item The cross section for single nucleon (e,e$'$p) knockout is only
40\% to 60\% of that predicted by Distorted Wave Impulse Approximation
(DWIA) analysis assuming four p-shell and two s-shell protons.  This
is consistent with the Saclay results and all other published
quasielastic data.  In the delta-region measurements, as expected, the
single-nucleon-knockout is virtually invisible.
\item In stark contrast to the transverse response function, the
longitudinal response function measured at $q=400$ MeV/c 
is consistent with zero for $E_m \ge 50$ MeV.  This suggests that
single nucleon knockout is minimal beyond $E_m = 50$ MeV.
\item A considerable fraction of the cross section occurs at $E_m >
50$ MeV.  The separated measurement at $q=400$ MeV/c indicates that
this strength is transverse and begins at $E_m \approx 27$
MeV, the threshold for 2-nucleon emission.  This ``continuum''
strength is attributed to two- and multi-nucleon knockout. 
The continuum strength persists in the measurements on the delta peak,
and constitutes a large fraction of the total cross section even where
pion production is expected to dominate.  Note that excess transverse
cross section was observed on other nuclei at missing energies above
the 2-nucleon emission threshold.\cite{Lanen}
\item No abrupt change in cross section was seen at pion threshold,
$E_m \approx 155$ MeV, for $q = 775$ MeV/c, the only quasielastic
measurement so far to probe sufficiently high missing energies.
However, an abrupt increase in the cross section was seen in the
delta-region measurements.
\end{enumerate}
Figure~\ref{q:omega} shows the momentum and energy transfer regions of
the quasielastic, dip, and $\Delta$ measurements at Bates, including this
experiment.

Kester {\em et al.}\cite{Kester} have recently measured the
$^{12}$C(e,e$'$p) reaction in the dip region at a variety of angles
away from parallel kinematics.  They find that large-angle cross
sections can be explained by meson-exchange-currents and intermediate
deltas, while smaller-angle cross sections suggest correlated pair
emission.

\section{The Experiment}
We report two measurements of the $^{12}$C(e,e$'$p) reaction, at $q =
970$ and 990~MeV/c.  Both were done in parallel kinematics.  The
energy transfers were respectively $\omega = 330$ and 475~MeV.  The
latter point is at the maximum of the C(e,e$'$) quasielastic peak, and
extends the investigation of the momentum-transfer dependence of the
C(e,e$'$p) reaction cross section measured at $q = 400$, 585, 775, and
827 MeV/c.  With both measurements, we investigate how the
single-nucleon and continuum cross sections depend on the energy
transfer on and below quasielastic kinematics.  The specific
kinematics are shown in table~\ref{kinematics} and figure~\ref{q:omega}.

We performed the experiment at the MIT-Bates Linear Accelerator Center
in Middleton, Massachusetts.  The recirculated electron beam had an
average energy of 696~MeV $\pm$~3~MeV for the $\omega=330$ MeV
measurement, and 796~MeV $\pm$~3~MeV for the $\omega=475$ MeV
measurement.  The beam had a duty factor of approximately 1\%, with
1--20 $\mu$A average (0.1--2 mA peak) current.  We used several
natural carbon targets with areal density or thickness ranging from 24
mg/cm$^2$ to 410 mg/cm$^2$.  We also used a spinning polyethylene
target to measure the elastic H(e,e$'$) reaction for normalization,
and tantalum and beryllium oxide targets for testing and calibration.

We used the magnetic spectrometers MEPS to detect electrons and OHIPS
to detect protons.  The polarity of OHIPS was reversed to detect
electrons during calibration measurements.  The
spectrometers are described in detail elsewhere\cite{Morrison}. In
each spectrometer, a scintillator array detected a particle passing
through the spectrometer's focal plane and triggered the readout
system.  A two-plane vertical drift chamber measured the particle's
trajectory at the focal plane.  MEPS used an Aerogel \v{C}erenkov
counter with an index of refraction of 1.05 to distinguish between
electrons and pions.

We identified coincidence events by the time elapsed between the
electron trigger in MEPS and the proton trigger in OHIPS.  The
coincidence time resolution was approximately 2~ns FWHM.  Accidental
events under the timing peak were subtracted, and this subtraction is
included in the statistical errors of the spectra.

\subsection{Calibrations, Corrections and Efficiencies}
We measured H(e,e) in MEPS, elastic C(e,e) in OHIPS and coincidence
H(e,ep) at various spectrometer magnetic fields to determine the
spectrometer constants and beam energies.  The uncertainties are 3 MeV
in the beam energy.

We calculated correction factors to account for losses due to many
effects including software track reconstruction, simultaneous events
in a wire chamber, more than one event per beam burst, and other
software and hardware limitations.  The correction factors varied from
run to run, ranging from 1.40 to 1.90.  Some correction factors were
deduced from run-to-run variations and are only valid up to an overall
normalization, discussed in the following section.

Because the (e,$\pi^-$p) cross section is much larger than the
(e,e$'$p) cross section at deep missing energies, we needed to reject
pions.  We used the $n=1.05$ Aerogel \v{C}erenkov counter in MEPS for
this purpose.  Electrons passing through the aerogel radiated
\v{C}erenkov light, whereas pions with momentum less than 430 MeV/c
did not radiate.  The electron detection efficiency of the Aerogel
\v{C}erenkov counter varied strongly with the MEPS magnetic field.
For $\omega=475$ MeV, the electron detection efficiency was 93\% and
the pion rejection efficiency was 99.5\%.  For $\omega=330$ MeV, the
electron detection efficiency was only 60\% and the pion rejection
efficiency was 98.5\%.  We also determined the electron detection
efficiency as a function of focal plane position.

To obtain the relative acceptance (including detection efficiency) of
the spectrometers as a function of focal plane position (ie: of
relative momentum), we measured the quasielastic C(e,e$'$) cross
section in MEPS and the C(e,p) cross section in OHIPS.  We varied the
magnetic field, placing particles with a given momentum at different
positions in the focal plane.  We deconvoluted the acceptance from the
single arm cross section to obtain the focal plane acceptance as a
function of relative momentum.  We then combined this with the
variation in \v Cerenkov counter electron detection efficiency with
focal plane position to get the total spectrometer relative
efficiency-acceptance product (hereafter called `relative acceptance').
We applied these relative acceptances to all of our data.  The
absolute normalization of the spectrometers is discussed in the next section.

\subsection{Normalizations}
To normalize the experiment absolutely, we measured the H(e,e$'$)
elastic cross section in MEPS, the H(e,e$'$p) elastic cross section
detecting electrons in MEPS and protons in OHIPS, and the C(e,e$'$)
elastic cross section in OHIPS.  We corrected these measured cross
sections for the relative acceptances as a function of
momentum (described in the previous section).  We then compared the
corrected measured H(e,e$'$p) cross section with Simon {\em et al.}'s
parametrization of the H(e,e$'$) cross section\cite{Mainz}, and the
corrected C(e,e$'$) cross section with the phase-shift calculation of
the program ELASTB\cite{Norum}.

Ideally, the H(e,e$'$p) measurement would fully normalize the
experiment after taking into account relative efficiencies and dead
times.  However, if the electron from H(e,e$'$p) enters MEPS,
kinematics restrict the proton to a small region within OHIPS's solid
angle.  C(e,e$'$p) protons populate the entire OHIPS solid
angle approximately uniformly.  Particles entering OHIPS near the
edges of OHIPS's collimator may not reach the focal plane.  These
losses affect the overall normalization, but H(e,e$'$p) alone would
not measure them.

We measured the elastic C(e,e$'$) cross section in OHIPS to
account for those losses, but the electrons from C(e,e$'$) did
not cover the OHIPS solid angle uniformly either.  At 17$^\circ$, the
C(e,e$'$) cross section is approximately inversely proportional
to the fourth power of the scattering angle.  Most electrons entered
OHIPS near the front of the angular acceptance.

We used the transport program TURTLE\cite{Carey}{} to model the
physical characteristics of OHIPS between the entrance near the target
and the focal plane, and to estimate the fraction of particles
entering the solid angle that reach the focal plane.  We used three
initial distributions of particles over the solid angle.  TURTLE gave
the following results for the indicated distribution of entering
particles:
\begin{itemize}
\item 100\% --- Uniform over the restricted H(e,e$'$p) region
\item 85\% --- Inversely proportional to $\theta^4$ as we expect for C(e,e$'$)
\item 89\% --- Uniform over the entire OHIPS solid angle as we expect for
		C(e,e$'$p)
\end{itemize}
The C(e,e$'$) cross section measured in OHIPS was (82 $\pm$
5)\% of the cross section calculated by ELASTB.  After applying the
correction functions calculated in the previous section for the
\v{C}erenkov counter inefficiency and the spectrometer acceptances as
a function of momentum, the H(e,e$'$) and H(e,e$'$p) measured cross
sections were the same, indicating that OHIPS had no additional losses. 
TURTLE's results were consistent with both.

The overall normalization factor is the product of the two terms:
\begin{itemize}
\item The Mainz H(e,e$'$p) cross section calculation divided by the
measured H(e,e$'$p) cross section --- 1.06 for $\omega = 330$~MeV, and
1.24 for $\omega = 475$~MeV
\item The OHIPS factor from TURTLE and C(e,e$'$), given by
$({1\over 0.89})({0.85\over 0.82\pm 0.05}) = 1.16\pm 0.07$.  The
factor of $({1\over 0.89})$ comes
from TURTLE for a uniformly illuminated solid angle.  The factor
$({0.85\over 0.82\pm 0.05})$ is a small correction to the TURTLE
normalization from the measured C(e,e$'$) cross section.
\end{itemize}
The normalization factors at the center of the focal plane (0\%
relative momentum) were 1.23 for $\omega = 330$~MeV, and 1.44
for $\omega = 475$~MeV.  Normalization factors at other locations on
the focal plane were the product of the focal plane center
normalization and the relative acceptance of the other location
determined as described in the previous section.

The systematic uncertainty in the C(e,e$'$p) cross section is
8\% for the entire missing energy spectrum, primarily due to beam
energy uncertainty coupled to the C(e,e$'$) and H(e,e$'$) cross
sections and statistical uncertainty in the normalization
measurements.  In addition, there is a further systematic uncertainty
of 4\% in the continuum region ($E_m > 50$ MeV) due to possible
residual pion contamination.

\subsection{Representation of the Differential Cross Section}
\label{Multipole}
We measured the coincidence cross section as a function of missing
energy for each of the two kinematics, at $\omega = 330$~MeV and
475~MeV, varying only the proton final momentum $p_f$.  For each
measurement, we represented the $\omega$ dependence of the cross
section within the $\omega$ acceptance of the electron spectrometer by
expanding the cross section around the central value of $\omega$ using
orthogonal polynomials:
\begin{equation}
{d^4\! \sigma \over d\Omega_e\, d\Omega_p\, d\omega\, dE_m} =
	\sum_{l=0}^{l_{max}} \alpha_l(E_m)
	P_l\left({\omega-\omega_0\over\Delta\omega/2}\right)
\label{LegendreExpansion}
\end{equation}
where $P_l(x)$ are Legendre polynomials, $\omega_0$ is the central
value, and $\Delta\omega$ is the width of the $\omega$ acceptance.
The experimental coefficients $\alpha_l(E_m)$ are determined from the
data using the method described in \cite{Morrison}.  For a given
$E_m$, the true differential cross section is expected to vary
smoothly with $\omega$, so $\alpha_l(E_m)$ should approach zero
rapidly as $l$ increases.  This expansion of the $\omega$-dependence
of the cross section is necessary since we lack sufficient
experimental statistics to determine a full two-dimensional
$(E_m,\omega)$ spectrum.

All $\alpha_l$ have the same units: picobarns/MeV$^2$-sr$^2$.
$\alpha_0(E_m)$ is an average of the cross section over the $\omega$
acceptance.  The nature of the average depends on the cutoff
$l_{max}$.  $\alpha_1(E_m)$ multiplies
$(\omega-\omega_0)/(\Delta\omega/2)$ in
equation~\ref{LegendreExpansion}; it measures the change of the cross
section over $\Delta\omega$. The ratio $\alpha_1/\alpha_0$, which
measures the relative change of the cross section with $\omega$, may
be more relevant in comparing the experiment with theory.  Higher
order terms ($\alpha_l$ with $l \ge 2$) multiply higher order
polynomials of $\omega$, and indicate the curvature of the cross
section.

The calculation of the coefficients $\alpha_l(E_m)$ depends somewhat
on the choice of cutoff $l_{max}$.  Values of $\alpha_l$ significantly
different from zero are available from the data for $l = 0$, 1, 2, and
3, although $\alpha_0$ and $\alpha_1$ yield the dominant features.  We
verified that $\alpha_l$ (for $l \leq l_{max}$) was roughly
independent of $l_{max}$ for $l_{max}=2$, 3, or 4.  $\alpha_0$
calculated using $l_{max}=0$ and using $l_{max}$ = 2, 3, and 4 differ
by less than 15\%.  For $l_{max}=0$, $\alpha_0$ is the average of the
cross section over the $\omega$ acceptance.  As $l_{max}$ increases,
the variation of the cross section over the $\omega$ acceptance is
described by the higher order terms so that $\alpha_0$ becomes the
cross section at the center of the $\omega$ acceptance.

The calculations we present use $l_{max} = 0$ and 3.  The cross
sections of the previous experiments at $q=400$, 585, and 775, and 827
MeV/c were averaged over the $\omega$ acceptance, corresponding to
$\alpha_0$ with $l_{max} = 0$.  Therefore, comparisons with previous
measurements use the results from $l_{max}=0$.

\subsection{Radiative Corrections}
We used the prescription of Borie and Drechsel\cite{Borie} to subtract
the radiative tails of the p-shell and s-shell peaks from the missing
energy spectra.  Computing these tails requires knowledge of the
coincidence cross section for all values of $\omega$ and $E_m$ less
than the experimental values.  Lacking this knowledge, we
calculated both the peak and radiative tail cross sections using the
Plane Wave Impulse Approximation (PWIA) and harmonic oscillator
initial state wave functions.  We scaled the tail calculation by the
ratio of the measured peak cross section to the calculated peak cross
section before subtracting the tail from the spectrum.

We calculated the Schwinger correction\cite{Schwinger,Maximon}, with a
hard photon cutoff of 11.5 MeV.  We multiplied the p-shell peak by the
Schwinger correction and subtracted the p-shell radiative tail from
the s-shell and continuum regions of the spectrum.  Then we multiplied
the s-shell peak (limited to $E_m = 50$ MeV) by the Schwinger
correction using the same cutoff and subtracted the s-shell tail from
the continuum region.  Finally we applied the Schwinger correction to
the continuum.  We did not attempt to calculate continuum tails as we
had no satisfactory model for them.

\section{Results and Discussion}
\subsection{Features of the Spectra}
Figures~\ref{spectra:high} and~\ref{spectra:low} show the Legendre
expansion of the radiatively corrected cross-section as a function of
missing energy ($\alpha_0$ through $\alpha_3$, calculated with
$l_{max}= 3$ [see section IIC for a description of the expansion]).
(Note the difference in scales among the plots.)  We see three
features in $\alpha_0$ for both kinematics:
\begin{itemize}
\item A peak centered at $E_m = 18$ MeV primarily due to single
nucleon knockout from the p-shell
\item A broader peak out to $E_m \approx 60$ MeV primarily due to
knockout from the s-shell, but with possible contribution from the
continuum.
\item Continuum strength at larger missing energy attributed to two-
and multi-nucleon knockout
\end{itemize}
Ulmer's $R_L/R_T$-separation at $q=400$ MeV/c\cite{Ulmer}
indicates that s-shell knockout becomes small at 50 MeV, and that the
continuum strength starts at 27 MeV.

We note that the ratio of s-shell to p-shell cross section is much
smaller at $\omega=330$ MeV than at $\omega=475$ MeV.  The continuum
strength ($E_m>50$ MeV) extends beyond $E_m = 300$ MeV for
$\omega=475$ MeV, but goes to zero at approximately $E_m = 90$ MeV for
$\omega=330$ MeV.  We do not see any increase in cross section at pion
threshold, $E_m \approx 155$ MeV.

The $\omega=475$ MeV $\alpha_0$ cross section spectrum appears to have
a peak around $E_m = 60$~MeV.  The peak does not appear in the
spectrum if we use a bin size of 6 MeV instead of the 3 MeV size used
in figure~\ref{spectra:high}, and we do not judge it statistically
significant.

The $\alpha_1$ spectra have features that correspond to the features
of the $\alpha_0$ spectra.  In the $\omega=330$ spectrum, there is a
narrow peak at 18 MeV and a broad peak beyond 25 MeV.  These have
corresponding peaks in the $\alpha_0$ spectrum, and indicate that the
cross section increases strongly across the $\omega$ acceptance.  The
continuum cross section beyond 50 MeV also has a large $\alpha_1$
relative to $\alpha_0$ indicating that it also increases strongly with
$\omega$.

In the $\omega=475$ $\alpha_1$ spectrum, the p-shell peak is small and
positive, indicating a small average increase in the cross section
over the $\omega$ acceptance.  The s-shell $\alpha_1$ is zero,
indicating that the cross section is on the average constant over the
$\omega$ acceptance.  At 60 MeV of missing energy, $\alpha_1$ becomes
positive, suggesting that the reaction mechanism has changed.  This is
consistent with the result of the L-T separation at $q=400$
MeV/c\cite{Ulmer} that s-shell single-nucleon knockout becomes small around 50
MeV.  Beyond 110 MeV in missing energy, $\alpha_1$ is consistent with
zero, indicating no $\omega$ dependence within the acceptance.

Although $\alpha_0$ and $\alpha_1$ exhibit the most dominant and
statistically significant features, $\alpha_2$ and $\alpha_3$ display
some features.  For $\omega=475$ MeV, $\alpha_2$ is consistent with
zero, but $\alpha_3$ has a statistically significant negative value in
the s-shell region and possibly in the p-shell region, indicating a
measurable curvature in the cross section as a function of $\omega$.
For $\omega=330$ MeV, $\alpha_2$ and $\alpha_3$ are consistent with
zero except in the p-shell region, where they are both negative.  We
offer no interpretation of $\alpha_2$ and $\alpha_3$ in this paper.

\subsection{Momentum Distributions}
\label{momentum:dists}
The $\alpha_0$ and $\alpha_1$ spectra for the p and s shells
collectively exhibit qualitative features consistent with the momentum
distributions expected of p- and s-shell orbitals, as displayed in
figure~\ref{mom:dist}.  The s-shell momentum distribution has its
maximum around zero missing momentum, while the p-shell momentum
distribution has its maxima around $\pm$100 MeV/c, and reaches a
minimum at zero.

In parallel kinematics, the energy transfer is related to the missing
momentum by
\begin{displaymath}
\omega - {Q^2 \over 2M} \approx {{\bf p} \cdot {\bf q} \over M} =
{p_m^\parallel q \over M}
\end{displaymath}
for quasielastic single-nucleon knockout.  Choosing $\omega$
determines the central value of the parallel component of the missing
momentum.  Although the experiment was centered at parallel
kinematics, its finite angular and momentum acceptances 
covered a large range of the missing momentum perpendicular
to $\vec q$.  The parallel and perpendicular components of the missing
momentum ranges sampled by the experiment are shown in figure ~\ref{mom:dist}.  
The central parallel missing momenta for the measurements are
given in table~\ref{kinematics}.  At $\omega=475$ MeV, the parallel
component of the missing momentum covers approximately $-30$ MeV $<
p_m^\parallel < 100$ MeV (see figure~\ref{mom:dist}).  It is greater for
the p-shell than for the s-shell, reflecting the difference in binding
energy.  The s-shell momentum distribution is near its maximum. Thus
the s-shell 
cross section should be flat in $\omega$ (ie: $\alpha_1$ should be
small).  The p-shell cross section should increase slightly with
$\omega$.  We see these features in the $\alpha_0$ and $\alpha_1$
spectra in figure~\ref{spectra:high}.

At $\omega=330$ MeV, the central parallel missing momentum is much larger than
$-100$ MeV/c.  The p shell should dominate and both the p- and s-shell
cross sections should increase strongly with $\omega$.  $\alpha_0$ and
$\alpha_1$ in figure~\ref{spectra:low} reflect these traits.  The
p-shell cross section is much larger relative to the s-shell at
$\omega=330$~MeV than at $\omega=475$~MeV.

\subsection{Distorted Wave Impulse Approximation}
\label{DWIA:section}
We compared the observed single-particle knockout strength from each
shell with factorized Distorted Wave Impulse Approximation (DWIA)
cross section calculations.  We integrated the observed cross section
over missing energy from 10 MeV to 27 MeV for the p-shell, and 27 MeV
to 50 MeV for the s-shell.  The factorized DWIA cross section is given
by
\begin{equation}
{d^4\!\sigma \over d\Omega_e\, d\Omega_p\, d\omega\, dE_m} = E_f\,p_f
\sigma_{ep} |\phi^D({\bf p}_m,{\bf p}_f)|^2 f(E_m)
\label{DWIA:crosssection}
\end{equation}
where $\sigma_{ep}$ is deForest's CC1 off-shell electron-proton cross
section\cite{deForest}; $f(E_m)$ is the missing energy distribution
for the shell, normalized to a unit area; and $|\phi^D({\bf p}_m,{\bf
p}_f)|^2$ is the effective distorted momentum distribution of the
shell.  We used a delta function for $f(E_m)$ to describe the p-shell,
and a quadratic function between 30 and 50 MeV to describe the
s-shell.

Giusti and Pacati\cite{GiustiPacati} have calculated the effects of
Coulomb distortions of the electron wave function.  They find effects
of approximately one to two percent for $^{12}$C in  parallel
kinematics at an electron energy of 350 MeV.  They also find that the
effects decrease with initial energy.  Since we performed this
experiment at higher energies, we can disregard electron distortions.

We calculated $|\phi^D({\bf p}_m,{\bf p}_f)|^2$ using the program
PEEPSO, based on the non-relativistic (e,e$'$p) formalism of
Boffi\cite{Boffi}.  PEEPSO converts the relativistic Dirac optical
potential into a Schr\"odinger-equivalent potential including
spin-orbit terms, and then solves the Schr\"odinger Equation and
calculates the unfactorized (e,e$'$p) cross section for each shell,
with a given separation energy, at the center of the spectrometer
solid angle acceptances.  The effective distorted momentum
distribution is this calculated cross section divided by $E_f\, p_f
\sigma_{ep}$.  We used Woods-Saxon proton wave functions as measured
by van der Steenhoven {\em et al.} at NIKHEF \cite{Steenhoven} for the
initial bound states.

The optical potentials are fit to C(p,p) elastic scattering
results for different proton energies. We used the optical
potential of Hama {\em et al.}\cite{Hama} for the $\omega=475$ MeV
measurement.  For the $\omega=330$ MeV point, we calculated cross
sections from the Hama potential and also from the parameterization of
Meyer {\em et al.}\cite{Meyer}.  The Meyer potential is only fit to
C(p,p) elastic scattering data for 200 to 300 MeV protons; we
extrapolated it using the parametrized expressions.

We substituted the momentum distribution derived from PEEPSO into the
factorized expression, equation~\ref{DWIA:crosssection}, to obtain the
cross section over the entire experimental solid angle and energy
ranges.  From this we derived theoretical predictions for
$\alpha_l(E_m)$ as described in section~\ref{Multipole}, averaged over
the solid angle acceptances, using $l_{max}$ equal to 0 and 3 in
equation~\ref{LegendreExpansion}.

Tables~\ref{DWIA:low} and~\ref{DWIA:high} display the results of the
calculations along with the data.  The data differ from the
calculations; the ratio is the `data-theory-ratio'.\footnote{Other
experiments refer to the `data-theory-ratio' as a `spectroscopic
factor' and use it to infer properties of the proton initial state
wavefunction.  The tremendous variation of the data-theory-ratio with
$\omega$ in this experiment casts doubt on the theory and precludes
our using the term `spectroscopic factor'.}  The Hama and Meyer
potentials give similar results for the $\omega=330$ MeV p-shell, but
less similar results for the s-shell.  We used the average of the two
results for the calculated s-shell cross section, and assigned half
the difference (10\%) as an uncertainty in all the DWIA calculations
due to the choice of potential.  All other differences between the
Hama and Meyer potentials were less than 10\%.  We also calculated the
DWIA cross sections using a delta-function s-shell distribution in
missing energy.  The difference between the delta-function s-shell
result and the quadratic s-shell result was 10\% for $\omega=330$ MeV
and 3\% for $\omega=475$ MeV.  This contributed to the overall
uncertainty in the s-shell DWIA calculations.  We tested the
factorization approximation by calculating the distorted momentum
distribution (see equation~\ref{DWIA:crosssection}) from the PEEPSO
unfactorized cross section at fixed $(E_m, p_m)$ at the center and at
the edges of the spectrometer angular acceptances.  These differed by
5\% for the $\omega=475$ MeV p-shell and by 1\% for the s-shell and
for both shells at $\omega=330$ MeV.  The overall uncertainties were
15\% for the $\omega=330$ s-shell calculation, and 11\% for the
$\omega=330$ p-shell, and both $\omega=475$ shells.

We obtained the `data-theory-ratio' for each shell by dividing the
measured cross section by the calculation.  We used the average of the
Hama and Meyer calculations for the $\omega=330$ MeV theory cross
section.  The `data-theory-ratio' calculated for $l_{max}=0$ and 3 are
given in table~\ref{spect:factors}.  We use $l_{max}=0$ to compare
with results from prior papers.  (See section IIC for a description of
the Legendre expansion of the cross-section.) Note that these
comparisons of data to DWIA calculations are limited to the range of
missing energies and missing momenta ($\Delta p_m \approx 200$ MeV/c)
sampled by the measurements.  No (e,e$'$p) experiment has measured the
entire three dimensional missing momentum distribution.

At the quasielastic kinematics, $\omega=475$ MeV, the
data-theory-ratios are 0.40 for both the p- and s-shells.
Figure~\ref{spect:factor:plot} shows these data-theory-ratios, along
with those from previous quasielastic and dip measurements. The
data-theory-ratios appear to be constant or perhaps decrease slightly
with momentum transfer.  The s-shell region ($27 < E_m < 50$ MeV) also
includes two-nucleon knockout; this greatly increases the
uncertainties of the s-shell data-theory-ratios.

For $\omega=330$ MeV the data-theory-ratios are 0.85 for the
p-shell and 1.0 for the s-shell, close to the naive expectation.
  The 3-vector momentum transfer of 970
MeV is approximately the same as for $\omega = 475$~MeV ($ q= 990$
MeV/c).  

The p-shell data-theory-ratio is approximately equal to the s-shell
data-theory-ratio for both data sets even though the ratio of p-shell
cross section to s-shell cross section increases by factor of four
between $\omega=475$ MeV and $\omega=330$ MeV.  This lends credence to
the model.

Ryckebusch has calculated C($\gamma$,N) and C(e,e$'$p) differential
cross sections from models that include two-nucleon
knockout\cite{Ryckebusch,Ryckebusch2,Ryckebusch3}.  His single-nucleon
knockout calculations include meson exchange currents, Delta currents,
and Mahaux's prescription for the missing energy spreading of the
s-shell.  For the data presented in this paper, Ryckebusch's s-shell
knockout calculations match the above results; he obtains the same
data-theory-ratios of 1 for $\omega=330$ MeV and 0.4 for
$\omega=475$ MeV.  This also lends credence to the models.

This variation in  data-theory-ratios  from quasielastic kinematics to
low-$\omega$ kinematics is qualitatively similar to that observed by
van der Steenhoven {\em et al.}\cite{Steenhoven} who also measured a
significantly larger ratio of data to DWIA at large negative missing
momenta ($\omega \le Q^2/2M$) than at positive missing momenta
($\omega \ge Q^2/2M$).  Bernheim\cite{Bernheim} obtained a similar
result.  

The model of the (e,e$'$p) cross section may have to be modified at
large negative missing momentum.  This is suggested from the
measurement of $\alpha_1$ at $\omega=330$ MeV in table~\ref{DWIA:low}.
The ratio $\alpha_1/\alpha_0$ is 1.5 times theory for
the p-shell, indicating that the cross section is much steeper in
$\omega$ or missing momentum than theory predicts.  The reverse is
true for the s-shell.

Penn {\em et al.}\cite{Penn} have measured the C(e,e$'$p)
cross section for a similar momentum transfer, but a lower $\omega$
and larger p-shell central missing momentum: $\omega = 235$ MeV and
$|{\bf p}_m| = 240$ MeV/c.  In figure~\ref{mom:dist}, that would be
farther to the left than the $\omega=330$ MeV measurement.  Penn
obtained a p-shell data-theory-ratio of $0.45 \pm 0.05$.  This is
similar to  our $\omega = 475$ MeV measurement, but different from
$\omega=330$ MeV.  However, the ratio $\alpha_1/\alpha_0$ at
$\omega=330$ MeV is 1.5 times the DWIA calculation in
table~\ref{DWIA:low}.  Thus the experimental cross section decreases
more rapidly with decreasing $\omega$ than theory predicts, leading us
to expect a lower data-theory-ratio  at lower $\omega$ using the
same model.

We recognize limitations in the available DWIA models.  In particular,
variations due to different optical potentials are already included in
our estimate of the uncertainty of the data-theory-ratios. In
addition, the code PEEPSO does not include relativistic dynamics.
However, the factor of two difference between the $\omega=330$ MeV and
the $\omega=475$ MeV data-theory-ratios remains a challenge for
nuclear theory.

\subsection{Quasielastic C(e,e$'$) Cross Section}
We have also measured the single-arm quasielastic $^{12}$C(e,e$'$)
cross section for each energy transfer.  We used a model by Warren and
Weinstein\cite{Warren} to extrapolate the measured coincidence
single-proton-knockout cross section of each shell to the entire
$4\pi$ steradian nucleon solid angle.  We compared the sum of the p- and
s- shell extrapolations with the measured single arm cross section.
For $\omega=330$ MeV, the extrapolated coincidence cross section was
$0.93\pm0.04$ of the single arm cross section.  For $\omega=475$ MeV,
the extrapolated coincidence cross section was $0.50\pm0.05$ of the
single arm cross section.  These ratios are consistent with the
C(e,e$'$p) data-theory-ratios.

\subsection{Multinucleon Knockout and Other Processes}
In figure~\ref{spectra:high}, we see extensive cross section beyond
$E_m = 50$ MeV at quasielastic kinematics ($\omega=475$ MeV).  This
strength is approximately constant beyond about 100 MeV, and appears
to extend out to the deepest missing energy measured.  The strength is
similar to that seen in previous quasielastic measurements
\cite{Ulmer,Weinstein,Lourie}.  Below the quasielastic peak, at
$\omega=330$ MeV, the continuum strength is present, but far weaker
relative to the single-nucleon cross section, and is consistent with
zero beyond $E_m = 90$
MeV.  We plot the ratio of the multi-nucleon-knockout cross section
(integrated over $E_m > 50$ MeV) to the single-nucleon-knockout cross
section (integrated over $E_m < 50$ MeV) for various continuum regions
from previous experiments and the $\omega=475$~MeV measurement in
figure~\ref{multi:nucleon:ratios}.

We estimated the contribution of multi-step processes, such as
(e,e$'$N) followed by (N,p), to the continuum cross section, by
convoluting the PWIA nucleon knockout reaction with two models of
(N,p) scattering.  The first model uses the intra-nuclear cascade code
MECC-7\cite{MECC7} to monte carlo the propagation of nucleons through
the nucleus as a series of independent collisions with other nucleons.
The code enforces the Pauli exclusion principle in the collisions.
The second model uses C(p,p$'$) data at 300 MeV and 20$^\circ$, and at
500 MeV and 16$^\circ$\cite{Segal}.  We multiplied the results from
the C(p,p$'$) data by 1.5 to approximately include neutrons, because
the (e,e$'$N) cross section is approximately proportional to the square of the
magnetic moment, and $(\mu_n/\mu_p)^2 \approx 0.5$.  The results are
given in table~\ref{MECC7}, along with the measured cross sections
from this experiment.  These calculations can only account for less
than 6\% of the data beyond $E_m = 27$ MeV.  The MECC-7 calculation
produces almost no cross section beyond $E_m = 100$ MeV.  The
C(p,p$'$) based calculation reaches its maximum at $E_m = 70$--80 MeV,
but has a long tail reaching to the deepest missing energy.  Half its
cross section may lie beyond $E_m=100$ MeV.

The cross section out to 90 MeV in missing energy in both $\omega=330$
and $\omega=475$ MeV measurements has the approximate shape expected
from Takaki's model of two-nucleon knockout\cite{Takaki}.  However,
its magnitude is larger by a factor of 16\cite{Takaki2}.  Beyond 90
MeV, the shape at $\omega=475$ MeV is consistent with Takaki's
three-nucleon knockout model.  At $\omega=330$ MeV, there is no
strength beyond 90 MeV; the continuum strength up to 90 MeV should be
mostly due to two-nucleon knockout.

Both rescattering calculations (MECC-7 and C(p,p$'$)) and Takaki's
calculation used harmonic oscillator initial-state momentum
distributions.  It is unlikely that using bound states derived from
realistic Woods-Saxon potentials will change this
result at $\omega=475$ MeV where the initial momenta involved are
small.  Even at $\omega=330$ MeV, the initial momenta of 100 to 250
MeV/c are reasonably small.  In addition, the strong decrease of the
continuum cross section at large $E_m$ for $\omega=330$ MeV compared
to $\omega=475$ MeV indicates that an initial momentum distribution
plus rescattering cannot explain the continuum cross sections.
However, initial-state correlations could contribute to the cross
section at deep missing energy, because two nucleons share the
transfered energy and we detect only one nucleon.  The C(p,p$'$)
rescattering calculation shows a larger tail than MECC-7 calculation;
this may reflect such correlations.  If so, those correlations are not
strong enough to explain our continuum cross section when they are
part of the rescattering picture.

However, neither the C(p,p$'$) nor the MECC-7 calculations included
such correlations in the initial (e,e$'$N) reaction; the initial
nucleon bound state was a simple harmonic oscillator.  If the large
yield we see at deep missing energy results from strong initial-state
correlations, this is very interesting.  But this is unlikely to
explain the longitudinal response at $q=400$ MeV/c\cite{Ulmer} which
is small beyond $E_m = 50$ MeV.  The dynamical correlations should
influence both the longitudinal and transverse responses.

\begin{figure}
\vspace{1in}
\caption{Cross Sections calculated by
Ryckebusch\protect\cite{Ryckebusch3}.
The points are the measured cross section ($l_{max}=0$); the dot-dash
line is single-nucleon knockout from the s-shell; the dotted line (too
small too see in $\omega=475$ MeV) is from (e,e$'$pp); the dashed line
is from (e,e$'$pn); and the solid line is the total multinucleon
knockout cross section.  The cross section is displayed for $E_m > 25$
MeV, omitting the p-shell.}
\label{Ryckebusch:figure}
\end{figure}
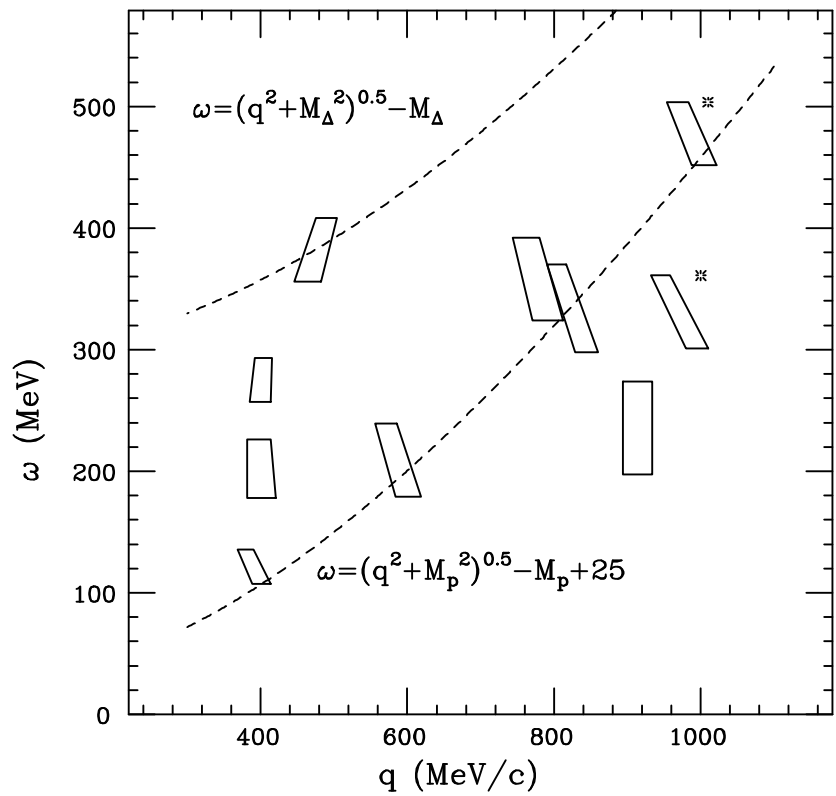
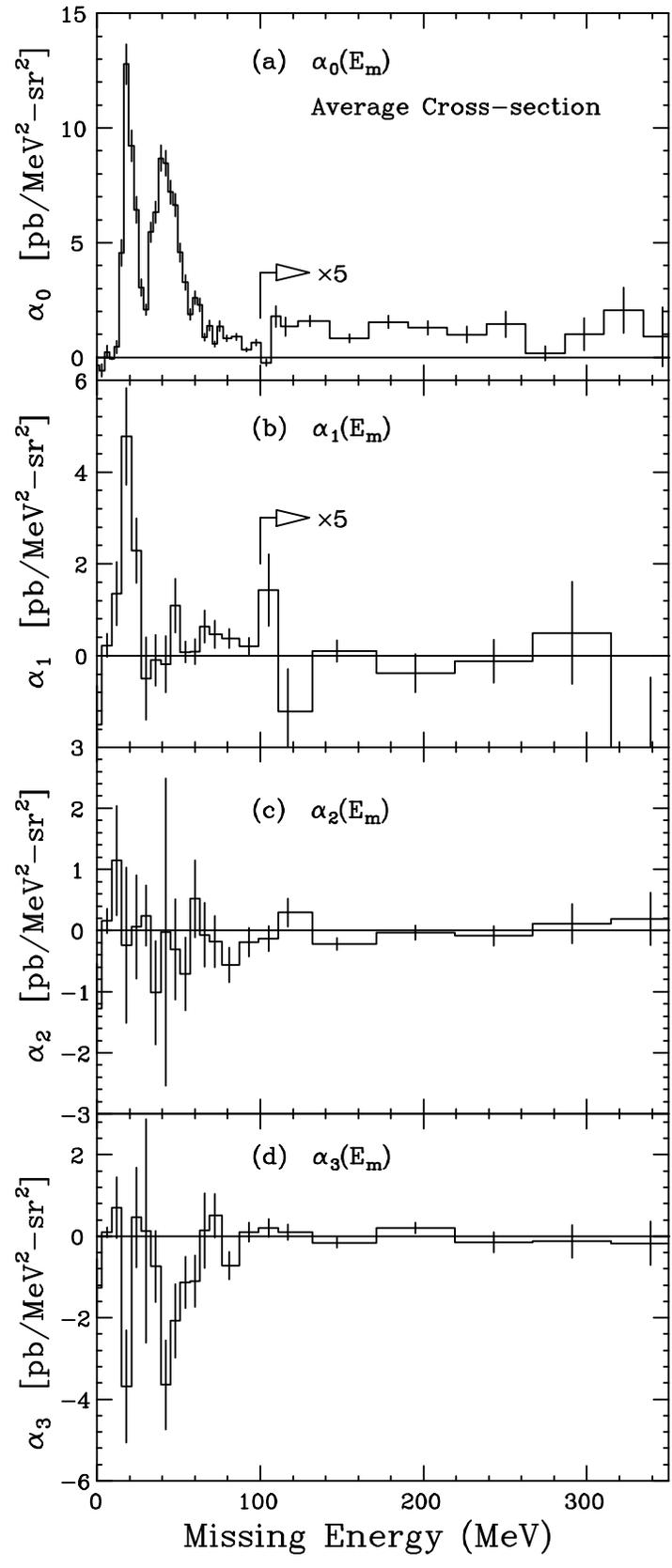
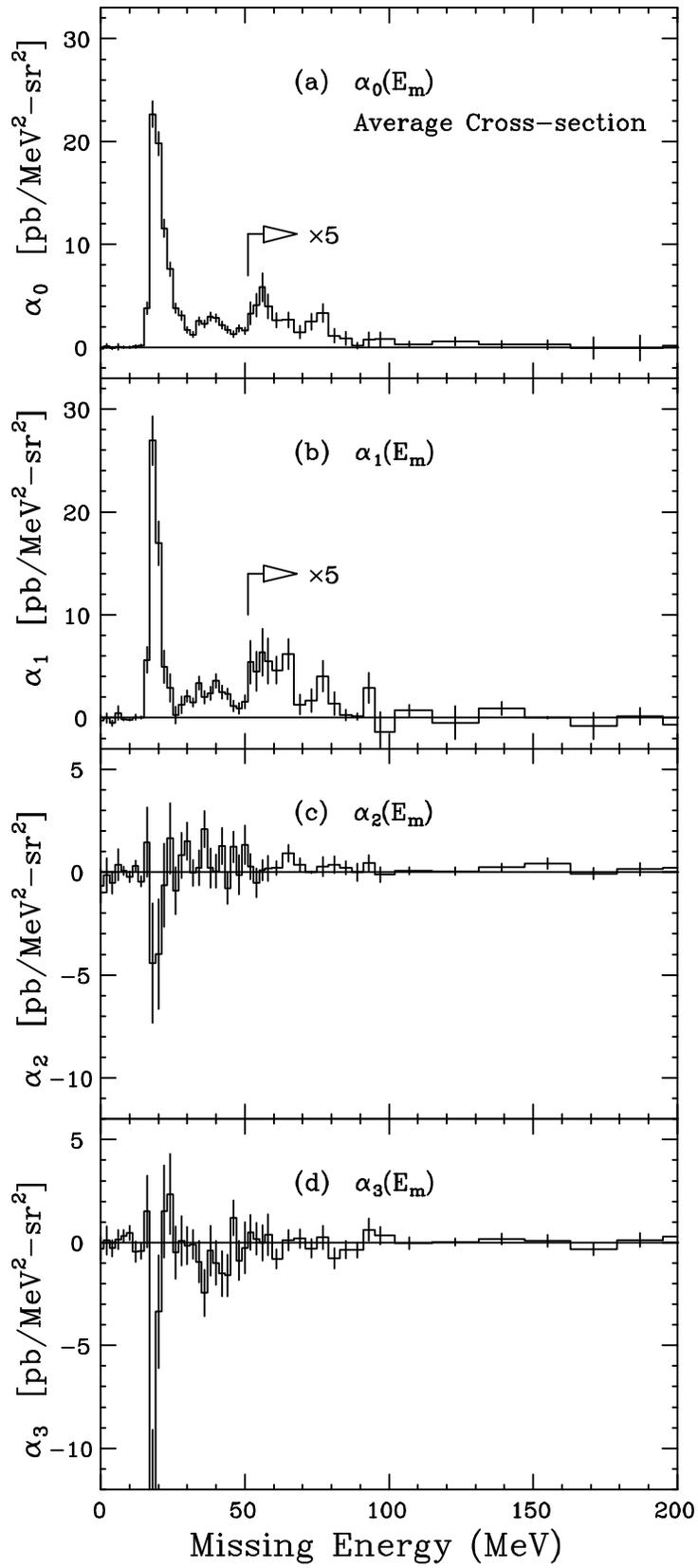
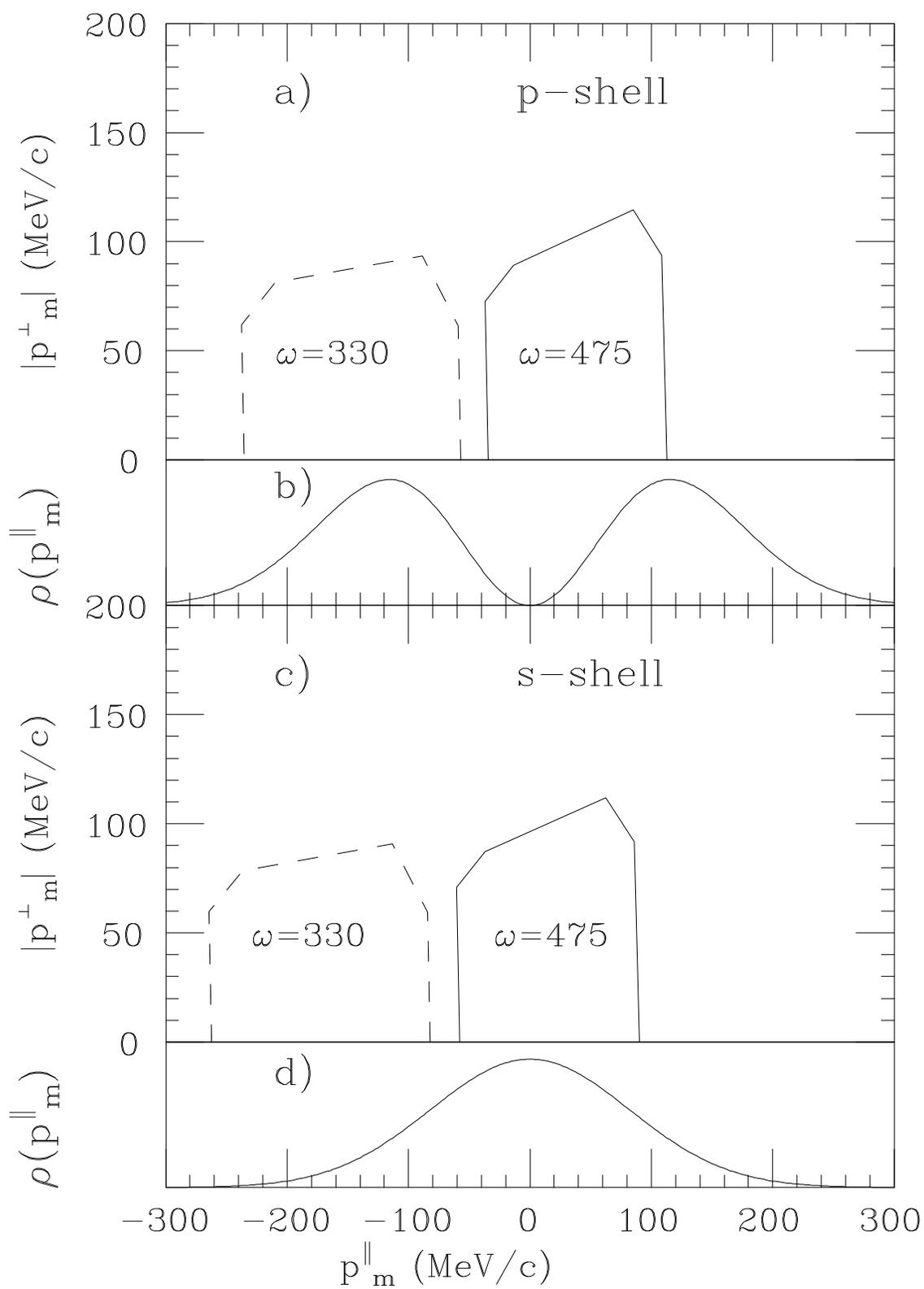
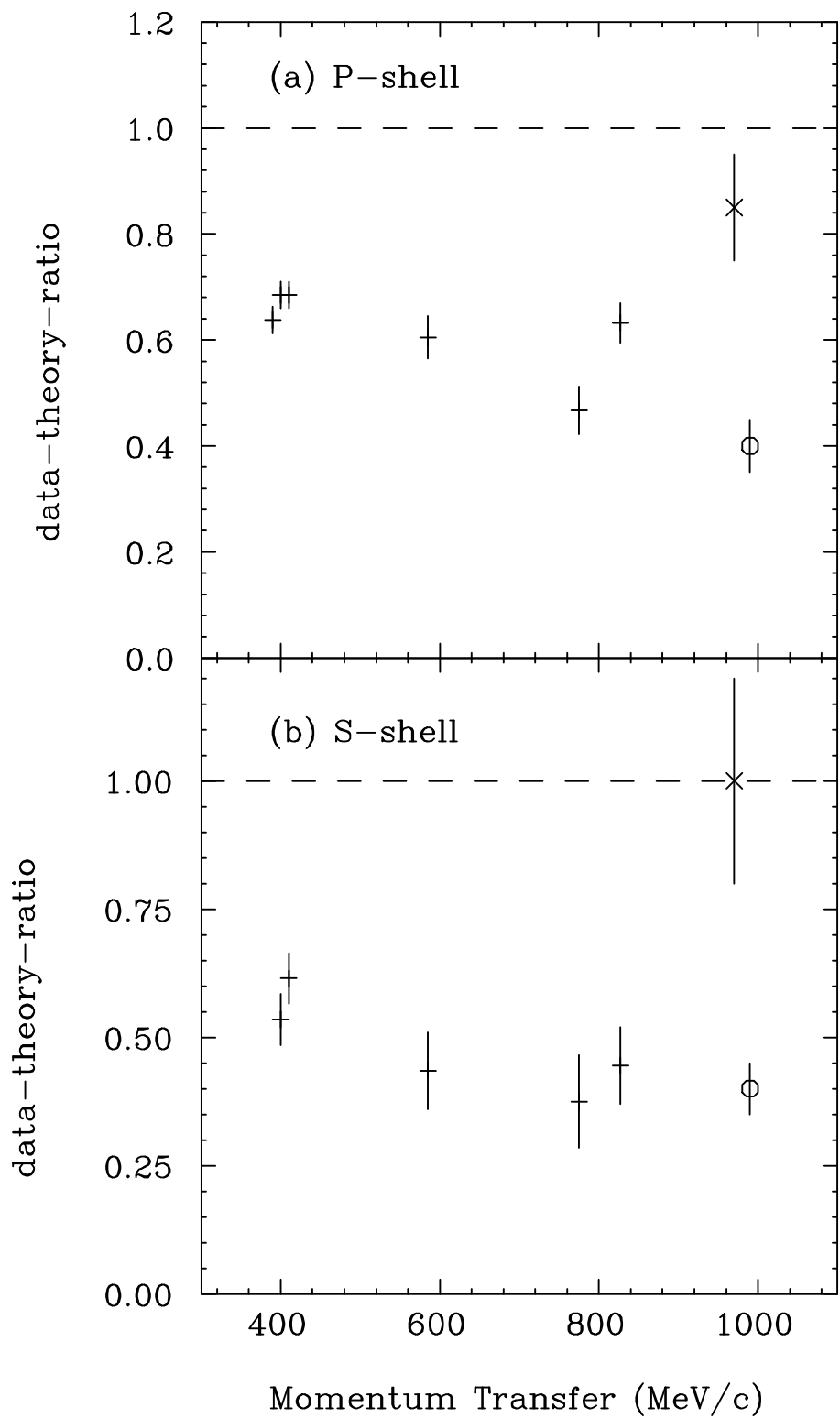
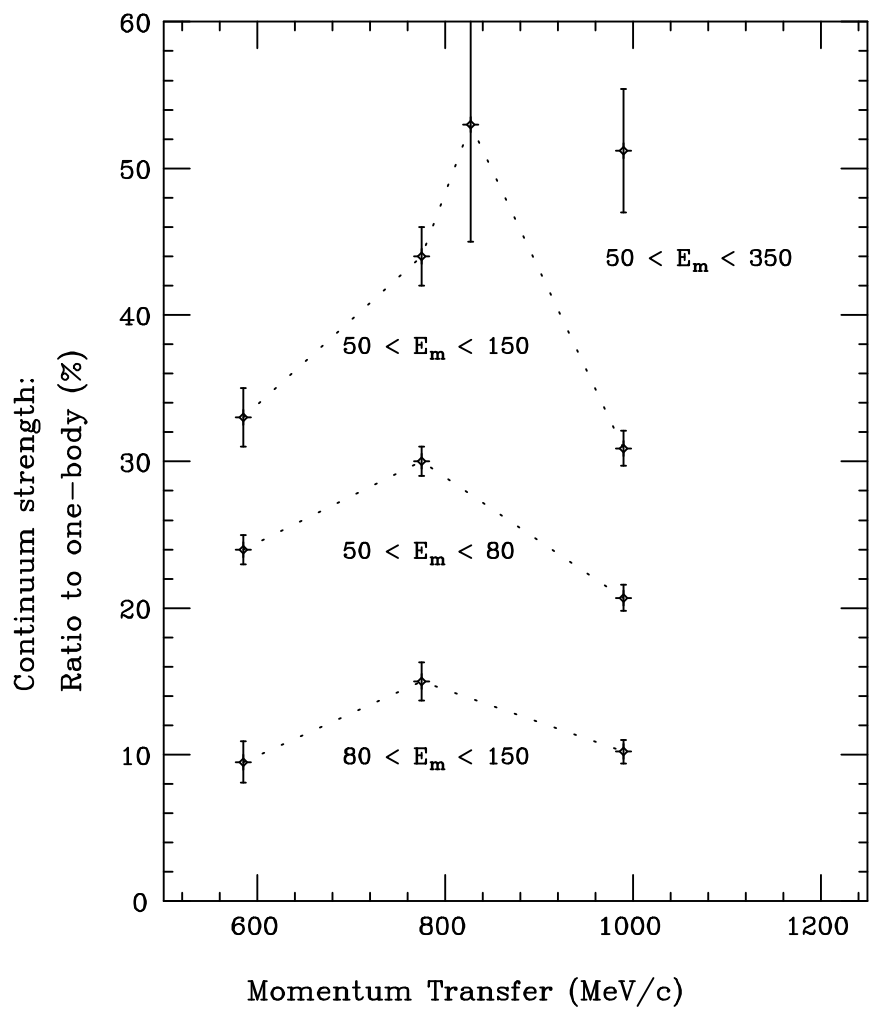
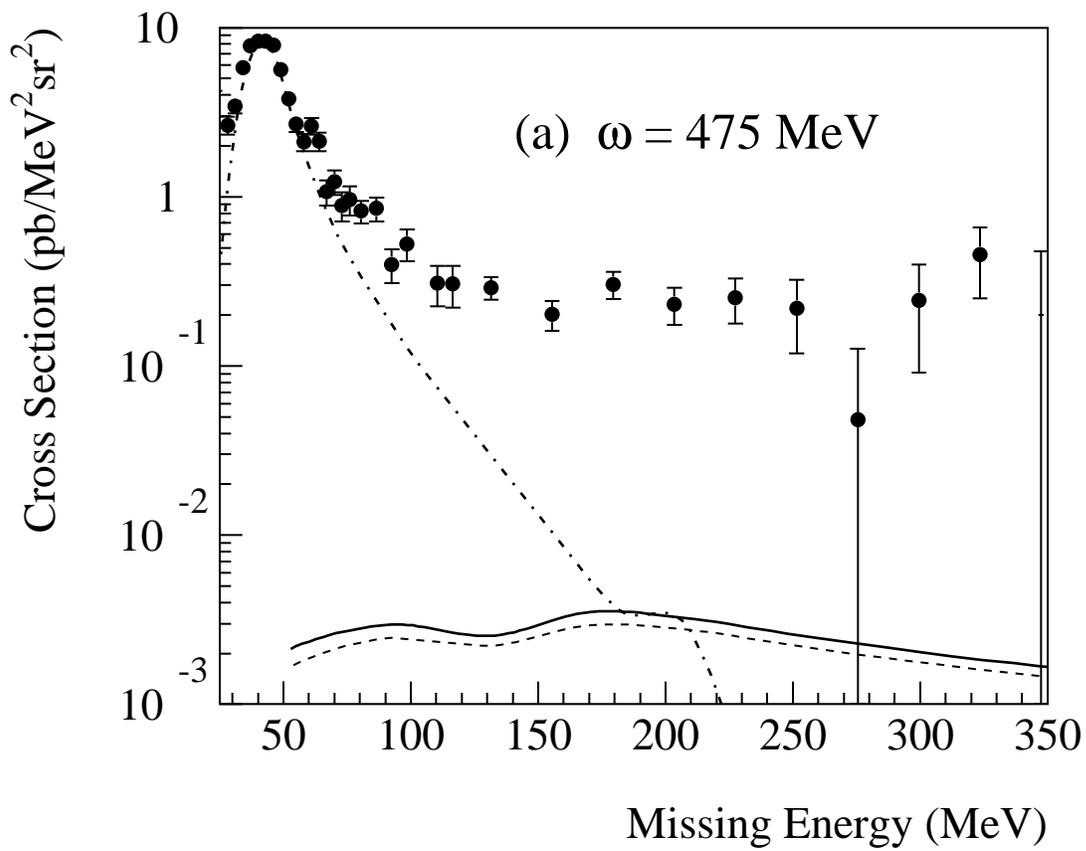
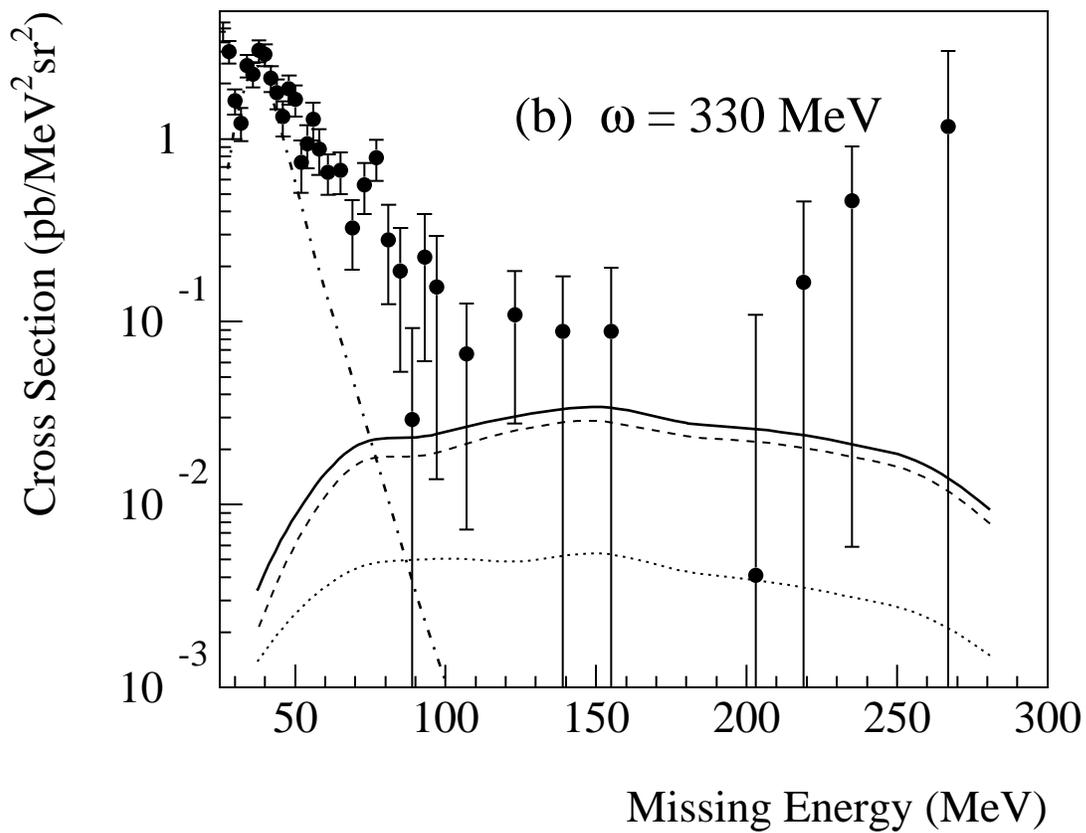

Later in section~\ref{Ryckebusch:Section} of this paper
(figure~\ref{Ryckebusch:figure}), we discuss calculations by
Ryckebusch using initial state Jastrow correlations.  Ryckebusch was
unable to generate more than one percent of our $\omega=475$~MeV
continuum cross section from the correlations.  Furthermore, one could
use Ryckebusch's missing energy spectrum as an input to a rescattering
calculation.  Ryckebusch's calculated s-shell (which does not include
correlations) fits our data after renormalization for
data-theory-ratios; it should therefore generate a rescattering cross
section comparable to our estimates.  Ryckebusch's continuum cross
section (which includes correlations) is $10^{-3}$ of his s-shell
cross section and $10^{-2}$ of our measured continuum cross section.
Thus, his continuum cross section cannot generate through rescattering
a cross section comparable to our data.

We see no increase in strength at pion threshold, $E_m \approx 155$
MeV.  Baghaei's PWIA $\Delta$-resonance pion-production
calculation\cite{Baghaei2,Baghaei} predicts more strength than we see
beyond pion threshold.  A calculation that we performed based on
Nozawa and Lee's pion-production model\cite{Nozawa} involving both
non-resonant and resonant production underpredicts the cross section
in that region by about half.  The calculation also predicts the
pion-production cross section to increase with $\omega$, resulting in
a positive $\alpha_1$.  Basic considerations of pion production
occurring at the tail of the $\Delta$ resonance also lead to the same
conclusion.  The measured $\alpha_1$ and the ratio $\alpha_1/\alpha_0$
are consistent with zero and inconsistent with the pion-production
prediction.  The results of the pion-production calculations are
presented in table~\ref{pion}.

We estimate an upper bound on the amount of two-nucleon knockout due
to $N-\Delta$ interactions.  Pion scattering experiments indicate that
the two-nucleon knockout cross section from the reaction
$N\Delta\rightarrow 2N$ is comparable to the pion-nucleon production
cross section due to $\Delta\rightarrow N\pi$\cite{NDelta}.  The
latter has to be less than the total integrated cross section above
$E_m=155$ MeV.  In the $\omega=475$ MeV measurement, if we assume that
the cross section for $N\Delta \rightarrow NN$ is less than or equal
to the integral of the experimental cross section for $E_m > 155$ MeV
and we distribute this strength in missing energy according to
Takaki's shape for two nucleon knockout in the region $50 < E_m < 150$
MeV, then $\Delta N \rightarrow NN$ can account for at most one-sixth
of the cross section for $50< E_m < 100$ MeV and none of the
cross section above 100 MeV.  At $\omega=330$ MeV, this can account
for none of the cross section.  However, one must be cautious; at
quasielastic kinematics, many of the $\Delta$s may not have enough
mass to decay into a real pion and a real nucleon.  The two-nucleon
cross section due to $N\Delta$ interactions could be greater than the
above estimate.

\subsection{Recent Multinucleon Calculations}
\label{Ryckebusch:Section}
Ryckebusch has calculated C($\gamma$,N) and C(e,e$'$p)
differential cross sections from models that include two-nucleon
knockout\cite{Ryckebusch,Ryckebusch2,Ryckebusch3}.  His single-nucleon
knockout calculations include meson exchange currents, Delta currents,
and Mahaux's prescription for the missing energy spreading of the
s-shell.  His two-nucleon knockout cross sections include Jastrow
correlations in addition.

These calculations fit the shape of the single nucleon knockout part of our
data.  Using Mahaux's s-shell spreading, these calculations also fit our data
out to $E_m \approx 60$ MeV.  This is consistent with the experiment
reported by Makins \cite{Makins} at $Q^2 = 1$ (GeV/c)$^2$.  Their
calculations appear to match their data using only
single-nucleon-knockout and radiative corrections, but their
cross section data extends only out to $E_m = 100$ MeV.  (Note that in
this paper we  use $E_m = 50$ MeV as the starting point for
multinucleon knockout since $R_L$ is small beyond that
point.)

Ryckebusch's calculations of real photon absorption understate the
measured C($\gamma$,N) cross sections at forward angles and at
high missing energies by about half\cite{Ryckebusch,Ryckebusch2}.  His
preliminary C(e,e$'$p) calculations\cite{Ryckebusch3} also
account for at most half the cross section beyond $E_m = 70$ MeV
measured in parallel kinematics at Bates for $q=585$ MeV/c,
$\omega=210$ MeV.  However, his calculations reproduce data taken in
non-parallel kinematics at NIKHEF\cite{Kester} far from quasielastic kinematics ---
$q=270$ MeV/c, $\omega=212$ MeV, and $\theta_{pq}=42^\circ$.

For the data presented in this paper, Ryckebusch's calculated
multinucleon knockout cross section is less than one percent of
the measured continuum cross section at $\omega=475$ MeV 
(see figure~\ref{Ryckebusch:figure}).  For
$\omega=330$ MeV, well below quasielastic kinematics, his calculations
are consistent with the measurement beyond $E_m = 100$ MeV, although
the measurement is also consistent with zero.  Ryckebusch
predicts more multinucleon knockout at $\omega=330$ MeV than at
$\omega=475$ MeV; we see the opposite effect.

Recently Benhar \cite{Benhar} calculated the continuum cross sections
at $E_m > 220$ using a correlated nuclear matter spectral function in
PWIA.  The magnitude of his calculated cross sections is consistent
with the data at $\omega = 475$ MeV and slightly overpredicts the data
at $\omega= 330$ MeV.  However, his calculated cross section decreases
much more rapidly with missing energy than does the data.  A
calculation using the $^{12}$C spectral function would be very
valuable to help us understand the large differences between the
$\omega =330$ and $475$ MeV measurements in both the valence knockout
and continuum regions.

\section{CONCLUSIONS}

The different data-theory-ratios  at $\omega = 330$~MeV and at
$\omega = 475$~MeV are consistent with the different cross sections
seen beyond $E_m = 50$ MeV.  At $\omega=330$~MeV, we see nearly four
p-shell and two s-shell protons, but little continuum cross section.
At $\omega=475$~MeV, we see half as many protons, but much more
continuum cross section, extending out to the deepest missing energy measured
(Figures~\ref{spectra:high} and~\ref{multi:nucleon:ratios}).  We
associate the cross section at $E_m > 50$ MeV with multinucleon
knockout.  We infer that some mechanism that increases with $\omega$
transforms some of the single-nucleon-knockout into
multinucleon-knockout.

The measurement at $\omega=475$ MeV strongly confirms prior
results that the (e,e$'$) reaction at quasielastic kinematics involves
strong many-body physics and reactions in addition to quasielastic
knockout.  These other reactions do not stem from either nucleon
rescattering or from $\Delta$ interactions.  

The $\omega=330$ MeV measurement indicates that well below
quasielastic kinematics, but above collective phenomena such as giant
resonances, the (e,e$'$) reaction is primarily single-nucleon
quasielastic knockout.  The data-theory-ratios, within large
uncertainties, are close to the expected values from the simple shell
model.  However there is still some residual many-body physics at that
low energy transfer.

These data, especially the strength at high missing energies, strongly
support the growing realization that the inclusive (e,e$'$)
quasielastic peak contains much more many-body physics than was
originally thought.  This additional complexity persists at large
momentum transfer and is not understood.  The low $\omega$ side of
the quasielastic peak appears to be dominated by the simple
single-nucleon knockout process, but some complexity still appears.

This work was supported in part by the Department of Energy under
contract \mbox{\#DE-AC02-76ERO3069}, and the National Science
Foundation.
%

%XXXXXXXXXXXXXXXXXXXXXXXXXXXXXXXXXXXXXXXXXXXXXXXXXXXXXXXXXXXXXXXXXXXXX

%
\onecolumn
\vskip 0pt

\begin{table}
\begin{center}
\begin{tabular}{|c|c|c|c|c|c|c|c|}
\hline
$E_0$   & $|\bf q|$     & $\omega$      & $\Delta\omega$ & $\theta_e$
	& $\theta_p $	& $p_m$ (s-shell)   & $p_m$ (p-shell) \\
(MeV)   &  (MeV/c)      & (MeV)         & (MeV)        & (Deg) 
        & (Deg)		& (MeV/c)	& (MeV/c) \\
\hline
696     & 970           & 330           & 65           & 129.7
        & 17.0		& -170.		&  -144. \\
796     & 990           & 475           & 60           & 118.1
	& 17.0		&   19.		&    43. \\
\hline
\end{tabular}
\end{center}
\caption{Experimental Kinematics --- Central Values}
\label{kinematics}
\end{table}
\begin{table}
\begin{center}
\begin{tabular}{|l|l|l|l|l|l|}
\hline & & $l_{MAX} = 0$ & \multicolumn{3}{c|}{$l_{MAX} = 3$} \\
\hline Shell & & $\alpha_0$ (pb/MeV--sr$^2$) & $\alpha_0$
	(pb/MeV--sr$^2$) &  $\alpha_1$ (pb/MeV--sr$^2$) &
	$\alpha_1/\alpha_0$ \\
\hline
P-shell & Data  & 130  $\pm$ 4 $\pm$ 10 & 139  $\pm$ 4 $\pm$ 11
		& 116 $\pm$ 8 & 0.83 $\pm$ 0.09 \\
	& Hama  & 38.5 $\pm$ 4 & 40.4 $\pm$ 4 & 20.2 $\pm$ 2
		& 0.50 $\pm$ 0.07 \\
	& Meyer & 37.7 $\pm$ 4 & 40.2 $\pm$ 4 & 23.8 $\pm$ 2
		& 0.59 $\pm$ 0.08 \\
\hline
S-shell & Data  & 50.6 $\pm$ 2 $\pm$  4 & 50.7 $\pm$ 2 $\pm$  4
		& 49.3$\pm$ 4 & 0.97 $\pm$ 0.12 \\
	& Hama  & 27.5 $\pm$ 4 & 27.3 $\pm$ 4 & 39.6 $\pm$ 6
		& 1.5  $\pm$ 0.3 \\
	& Meyer & 23.1 $\pm$ 4 & 22.9 $\pm$ 4 & 35.1 $\pm$ 6
		& 1.5  $\pm$ 0.4 \\
\hline
\end{tabular}
\end{center}
\caption{DWIA Calculations for $\omega=330$ MeV.
The data cross sections are integrated over the missing energy regions
$E_m < 27$ MeV for the P-shell, and 27 MeV $< E_m < 50$ MeV for the
S-shell.  The theory calculations are for one proton in the
appropriate shell.  The labels `Hama'\protect\cite{Hama} and
`Meyer'\protect\cite{Meyer} refer to the optical potentials used by
the DWIA calculations. $\alpha_0$ represents an average of the cross
section over the $\omega$ acceptance.  $\alpha_1$ represents how the
cross section increases over the acceptance.  See text for details.}
\label{DWIA:low}
\end{table}
\begin{table}
\begin{center}
\begin{tabular}{|l|l|l|l|l|}
\hline & & $l_{MAX} = 0$ & \multicolumn{2}{c|}{$l_{MAX} = 3$} \\
\hline Shell & & $\alpha_0$ (pb/MeV--sr$^2$) & $\alpha_0$
	(pb/MeV--sr$^2$) & $\alpha_1$ (pb/MeV--sr$^2$) \\
\hline
P-shell & Data & 92   $\pm$ 3 $\pm$ 7 & 100  $\pm$ 4 $\pm$ 8
		& 47 $\pm$ 8 \\
	& theory & 59.1 $\pm$ 7	& 70.1 $\pm$ 8 & 50.4 $\pm$ 6 \\
\hline
S-shell & Data & 150  $\pm$ 4 $\pm$ 12 & 144 $\pm$ 4 $\pm$ 12
		& 0 $\pm$ 14 \\
	& theory & 182  $\pm$ 20 & 180 $\pm$ 20 & -13.7 $\pm$ 20 \\
\hline
\end{tabular}
\end{center}
\caption{DWIA Calculations for $\omega=475$ MeV.
The data cross sections are integrated over the missing energy regions
$E_m < 27$ MeV for the P-shell, and 27 MeV $< E_m < 50$ MeV for the
S-shell.  The theory calculations are for one proton in the
appropriate shell.
$\alpha_0$ represents an average of the cross section 
over the $\omega$ acceptance.  $\alpha_1$ represents how the cross
section increases over the acceptance.  See text for details.}
\label{DWIA:high}
\end{table}
\begin{table}
\begin{center}
\begin{tabular}{|l|l|l|l|l|}
\hline &\multicolumn{2}{c|}{P-shell} &\multicolumn{2}{c|}{S-shell} \\
	& $l_{max} = 0 $ & $l_{max} = 3 $ & $l_{max} = 0 $
	& $l_{max} = 3 $ \\
\hline $\omega = 330$ MeV
	& 0.85 $\pm$ 0.11 & 0.86 $\pm$ 0.11
	& 1.00 $\pm$ 0.18 & 1.01 $\pm$ 0.18 \\
\hline $\omega = 475$ MeV
	& 0.39 $\pm$ 0.06 & 0.36 $\pm$ 0.05
	& 0.41 $\pm$ 0.06 & 0.40 $\pm$ 0.06 \\
\hline
\end{tabular}
\end{center}
\caption{Data-theory-ratios. 
The data-theory-ratios are the data cross sections divided by the DWIA
cross sections from tables \ref{DWIA:low} and \ref{DWIA:high}.  For
$\omega=330$ MeV, the average of the Hama and Meyer calculations was
used.}
\label{spect:factors}
\end{table}
\begin{table}
\begin{center}
\begin{tabular}{|l|l|l|}
\hline & $\omega=330$ MeV & $\omega=475$ MeV \\
\hline
Multiple Scattering with MECC-7
	& 2.4 pb/MeV-sr$^2$ & 9.6 pb/MeV-sr$^2$ \\
Multiple Scattering with C(p,p$'$) Data $\quad\times\quad 1.5$
	& 4.4 & 15.6 \\
Data S-shell C(e,e$'$p) ($E_m = 27$--50 MeV) & 51 $\pm$ 2
	& 150 $\pm$ 4 \\
Data Near-Continuum C(e,e$'$p) ($E_m = 50$--100 MeV)
	& 23 $\pm$ 2 & 68 $\pm$ 3 \\
Data Full Continuum C(e,e$'$p) ($E_m = 50$--350 MeV)
	& 23 $\pm$ 2 & 130 $\pm$ 10\\
\hline
\end{tabular}
\end{center}
\caption{Multiple-Scattering Cross Sections.
The measured cross sections, integrated over the given regions, are
compared with rescattering calculations convoluting (e,e$'$N) with
(N,p) cross sections based on MECC-7 calculations\protect\cite{MECC7} and
C(p,p$'$) data\protect\cite{Segal}.  The C(p,p$'$) cross sections results were
multiplied by 1.5 to approximately account for initial neutron
interactions.}
\label{MECC7}
\end{table}
\begin{table}
\begin{center}
\begin{tabular}{|l|l|l|l|}
\hline & $\alpha_0$ & $\alpha_1$ & $\alpha_1/\alpha_0$ \\
\hline Data & 29.2 $\pm$ 3.8 & -4.3 $\pm$ 6.2
	& -0.15 $\pm$ 0.21 \\
Baghaei ($\Delta$) & 45.3 & & \\
Lee (Born+$\Delta$) & 13.0 & 11.5 & 0.88 \\
\hline
\end{tabular}
\end{center}
\caption{Pion-Production Calculations.
The data is the $\omega=475$ MeV calculation integrated over $E_m >
155$ MeV, pion threshold.  Pion production calculations are based on
Baghaei\protect\cite{Baghaei} and Nozawa and Lee\protect\cite{Nozawa}.}
\label{pion}
\end{table}
\begin{figure}
\vspace{1in}
\caption{The $q$ and $\omega$ regions covered by the Bates $^{12}$C(e,e'p)
experiments \protect\cite{Ulmer,Weinstein,Lourie,Baghaei2,Penn}.  The
regions marked with an asterisk indicate the two measurements of this
paper.}
\label{q:omega}
\end{figure}
\begin{figure}
\vspace{1in}
\caption{Legendre expansion of the cross section vs missing energy for
$\omega=475$ MeV.  The quantities $\alpha_l(E_m)$ (with units
pb/MeV$^2$--sr$^2$) are coefficients in the expansion of the cross
section, equation~\ref{LegendreExpansion}.  $\alpha_0$ is an average
of the cross section over $\omega$; $\alpha_1$ is the linear change of
the cross section over the $\omega$ acceptance.  $\alpha_0$ and
$\alpha_1$ have been multiplied by 5 for $E_m > 100$ MeV for clarity.
$\alpha_2$ and $\alpha_3$ are the 2$^{nd}$ and 3$^{rd}$ order changes
in the cross section.}
\label{spectra:high}
\end{figure}
\begin{figure}
\vspace{1in}
\caption{Legendre expansion of the cross section vs missing energy for
$\omega=330$ MeV.  The quantities $\alpha_l(E_m)$ (with units
pb/MeV$^2$--sr$^2$) are coefficients in the expansion of the cross
section, equation~\ref{LegendreExpansion}.  $\alpha_0$ is an average
of the cross section over $\omega$; $\alpha_1$ is the linear change of
the cross section over the $\omega$ acceptance.  $\alpha_0$ and
$\alpha_1$ have been multiplied by 5 for $E_m > 50$ MeV for clarity.
$\alpha_2$ and $\alpha_3$ are the 2$^{nd}$ and 3$^{rd}$ order changes
in the cross section.}
\label{spectra:low}
\end{figure}
\begin{figure}
\vspace{1in}
\caption{Missing momentum acceptance of the experiment and schematic
momentum distributions.  a) P-shell experimental acceptances (the
magnitude of the perpendicular missing momentum $\vert \vec p_m^\perp
\vert$ vs. the parallel missing momentum $p_m^\parallel$) for the
$\omega =475$ MeV and $\omega=330$ MeV measurements; b) Qualitative
p-shell momentum distribution; c) same as 'a)' for the s-shell; d)
same as 'b)' for the s-shell.}
\label{mom:dist}
\end{figure}
\begin{figure}
\vspace{1in}
\caption{The data-theory-ratios from this and earlier experiments
in the p-shell (top plot) and the s-shell (bottom plot)
\protect\cite{Ulmer,Weinstein,Lourie,Baghaei2,Penn}.
The data-theory-ratio is given by the measured cross section
integrated over the peak in missing energy, divided by the DWIA
calculation.  The $\omega=330$ MeV data-theory-ratio for each shell is
identified by an $\times$; the $\omega=475$ MeV ratios are circles.
Previously published spectroscopic factors are divided by the naive
shell occupancy (p-shell=4, s-shell=2) to obtain data-theory-ratios.}
\label{spect:factor:plot}
\end{figure}
\begin{figure}
\vspace{1in}
\caption{The ratio of multinucleon knockout ($E_m > 50$ MeV) to single
nucleon knockout ($E_m < 50$ MeV) for this experiment ($\omega=475$
MeV) and earlier experiments
\protect\cite{Ulmer,Weinstein,Lourie,Baghaei2,Penn}}
\label{multi:nucleon:ratios}
\end{figure}
\end{document}